\documentclass[fleqn,12pt,twoside]{article}
\usepackage{espcrc1}

\usepackage{graphicx}
\usepackage[figuresright]{rotating}

\newcommand{\AmS}{{\protect\the\textfont2
  A\kern-.1667em\lower.5ex\hbox{M}\kern-.125emS}}

\title{Nucleon-nucleon momentum correlation function for light nuclei }

\author{Y. G. Ma\thanks{E-mail: ygma@sinap.ac.cn}, X. Z. Cai,
J. G. Chen, D. Q. Fang, W. Guo, G. H. Liu,  C. W. Ma, E. J. Ma,
 W. Q. Shen,  Y. Shi, Q.M. Su, W.D. Tian, H.W. Wang, K. Wang,
Y.B. Wei, T.Z. Yan
\address[SINAP]{Shanghai Institute of Applied Physics, Chinese
Academy of Sciences, P. O. Box 800-204, Shanghai 201800, China}}

\begin{document}

\maketitle

\begin{abstract}
Nucleon-nucleon momentum correlation function have been  presented
for nuclear reactions with neutron-rich or proton-rich projectiles
using a nuclear transport theory, namely Isospin-Dependent Quantum
Molecular Dynamics model. The relationship between the binding
energy of projectiles and the strength of proton-neutron
correlation function at small relative momentum has been explored,
while proton-proton correlation function shows its sensitivity to
the proton density distribution. Those results show that
nucleon-nucleon correlation function is useful to reflect some
features of the neutron- or proton-halo nuclei and therefore
provide a potential tool for the studies of radioactive beam
physics.
\end{abstract}
¡¡¡¡

\vspace{1cm}

Both the enhancement of total reaction cross section of light
nuclei induced reaction and the narrowing of the momentum
distribution of the projectile-fragments care seen as the possible
evidences of the halo nuclei \cite{Tanihata,Kobayashi}. In
addition, the weakening of the neutron-neutron momentum
correlation function was also reported for the halo-nuclei induced
system \cite{Ieki,Marques1}.  Since the nucleon-nucleon momentum
correlation function is related to the geometrical and kinematic
information \cite{Pratt}, i.e. spacial-time information, which
gives the fact that the large spacial separation will give a weak
correlation function. Therefore, the extended neutron or proton
density distribution of halo nuclei could reveal weaker
neutron-neutron or proton-proton correlation function.

In this work, we shall discuss some features of  the
nucleon-nucleon momentum correlation function in light nuclei
induced reactions in the framework of quantum molecular dynamics
(QMD) model  \cite{Aichelin}. Firstly, we would like to recall the
momentum correlation function technique. Experimentally, the
correlation function is defined as the ratio between the measured
two-particle distribution and the product of the independent
single-particle distributions,  $
              C(p_1,p_2)  = \frac{dn^2/dp_1dp_2}{dn/dp_1 dn/dp_2},
$where $d^2n/dp_1dp_2$ represents the correlated two-particle
distribution and $dn/dp_1$ and $dn/dp_2$ is the independent
single-particle distribution of particle 1 and 2, respectively.
While in the model calculation as we will present in this work,
the standard Koonin-Pratt formalism \cite{Pratt} was used to
construct the two-particle correlation function by convoluting the
emission function $g(\mathbf{p},x)$, i.e., the probability for
emitting a particle with momentum $\mathbf{p}$ from the space-time
point $x=(\mathbf{r},t)$, with the relative wave function of the
two particles, i.e.,
\begin{equation}
C(\mathbf{P},\mathbf{q})=\frac{\int d^{4}x_{1}d^{4}x_{2}g(\mathbf{P}%
/2,x_{1})g(\mathbf{P}/2,x_{2})\left| \phi (\mathbf{q},\mathbf{r})\right| ^{2}%
}{\int d^{4}x_{1}g(\mathbf{P}/2,x_{1})\int
d^{4}x_{2}g(\mathbf{P}/2,x_{2})}. \label{CF}
\end{equation}%
where  $\mathbf{P(=\mathbf{p}_{1}+\mathbf{p}_{2})}$ and $\mathbf{q(=}%
\frac{1}{2}(\mathbf{\mathbf{p}_{1}-\mathbf{p}_{2}))}$ are the
total and relative momenta of the particle pair respectively, and
$\phi (\mathbf{q}, \mathbf{r})$ is the relative two-particle wave
function with $\mathbf{r}$
being their relative position, i.e., $\mathbf{r=(r}_{2}\mathbf{-r}_{1}%
\mathbf{)-}$ $\frac{1}{2}(\mathbf{\mathbf{v}_{1}+\mathbf{v}_{2})(}t_{2}-t_{1}%
\mathbf{)}$.

Before we use Koonin-Pratt formalism to construct the correlation
function, an event generator is necessary to get the phase space
information of emission nucleons. In the present work,
Isospin-dependent Quantum Molecular Dynamics (IDQMD) transport
model has been used to give the particle production and their
phase space information, and then construct a two-particle
correlation function.

The Quantum Molecular Dynamics (QMD) approach is an n-body theory
to describe heavy ion reactions from intermediate energy to 2 A
GeV. It includes several important parts: the initialization of
the target and the projectile nucleons, the propagation of
nucleons in the effective potential, the collisions between the
nucleons, the Pauli blocking effect and the numerical tests. A
general review about QMD model can be found in \cite{Aichelin}.
The IDQMD model is based on QMD model affiliating the isospin
factors, which includes the mean field, two-body nucleon-nucleon
(NN) collisions and Pauli blocking  etc \cite{Ma-hbt,Wei}.

We simulate the reaction of $^{11}Li$ fragments into $^{9}Li$ + 2n
at 28 MeV/u. The two-halo neutron correlation function and the
core neutron-neutron correlation functions have been calculated.
The right panel of Fig. 1 shows the experimental \cite{Ieki} and
our calculation correlation functions for neutron-pair. The two
halo neutrons in calculation are defined as the emitted neutrons
in coincidence with $^{9}Li$ core and the neutrons in the insert
of the figure are the ones formed  the core (i.e., $^9Li$) inside.
The solid line is the calculated two-halo-neutrons correlation
function and the dashed line in the insert is the core
neutron-neutron correlation function. A corresponding carton can
be found in the left panel of the figure.  It shows clearly that
the correlation function between the two halo neutrons reproduces
the experimental data fairly well but the one between the core
neutrons can not, which indicates that the source distributions of
the halo neutrons and the neutrons inside the core are totally
different, the former is much looser so that the two neutrons are
easier to be separated, which is consistent with the small two
neutron separation energy. This directly leads to the loose-bound
neutron density distribution of $^{11}Li$  which is the major
reason for the abnormal larger total reaction cross section and
the very narrow momentum distribution of the fragment $^{9}Li$
\cite{Tanihata,Kobayashi}. From the above discussion, it is clear
that the correlation strength at small relative momentum ($q$) is
sensitive to the spatial extent of the  emission source of
neutrons, some information about the halo configuration could be
learned from there.

\begin{figure}
\vspace{-0.6truein}
\begin{center}
\includegraphics[scale=0.45]{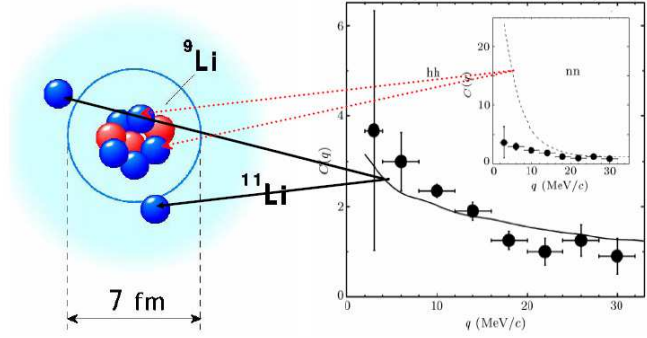}
\end{center}
\vspace{-0.4truein} \caption{\footnotesize Left panel: a carton of
the $^{11}$Li: core $^9Li$ plus two halo neutrons; the solid and
dashed lines point to the corresponding neutron-neutron
correlation function of the right panel;  Right panel: The solid
circles with error bars are the two-halo correlation functions in
the collision of $^{11}Li$ fragmented into $^{9}Li$ + 2n at 28
MeV/n \cite{Ieki}. The solid line is the calculated two-halo
neutrons correlation function and the dashed line in the insert is
the calculated two core neutrons correlation function.}
\label{Fig1}
\end{figure}

Based on the above success to fit the data, we further made a
prediction for the binding energy ($E_{b}$) dependence of n-p
correlation function strength for light isotopes. To this end,  we
select $^{12}C$  as the target and  $Li$, $C$ and $N$-isotopes as
the projectile, respectively. The reactions are simulated at
incident energy of 800 MeV/u and head-on collisions. The emitted
nucleons which are taken into account in the correlation function
calculation are all from the projectiles.

In order to make a quantitative comparison for the momentum
correlation function from  different isotopes, the strength of n-p
correlation function at very small relative momentum, $q$ = 5
MeV/c, has been extracted. Fig. 2 shows  the behavior of the
proton-neutron correlation function strength ($C_{PN}$) as a
function of binding energy of the projectiles. The fact that
$C_{PN}$ shows a rising with the increasing $E_{b}$ reflects in
some extent the larger $C_{PN}$ corresponds to a more compact
system, which is reasonable since  the mean compactness between
nucleons will change stronger with the increasing of the binding
energy, therefore $C_{PN}$ can more or less reveal the compactness
of the nuclei.
\begin{figure}
\begin{center}
\vspace{-0.6truein}
\includegraphics[scale=0.18]{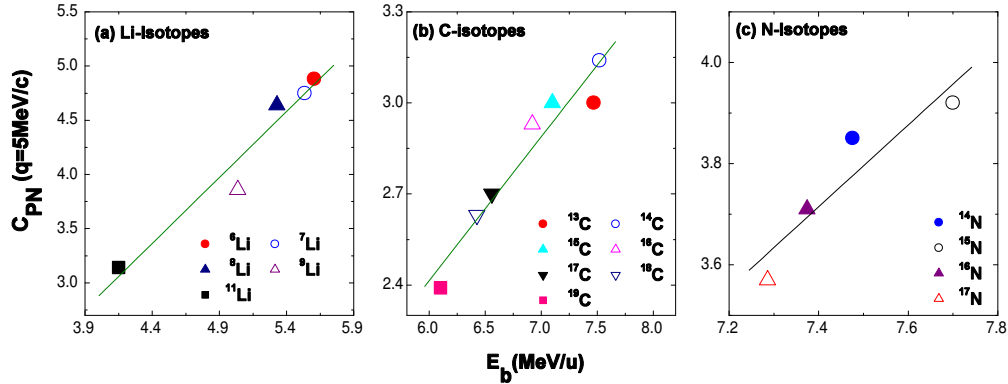}
\end{center}
\vspace{-0.9truein} \caption{\label{Fig2}  The relationship
between the strength of proton-neutron correlation function
$C_{PN}$ at 5 MeV/c and the binding energy per nucleon of the
projectiles for different isotopes:  $Li$ (a), $C$ (b) and $N$
(c). }
\end{figure}

\begin{figure}
\vspace{-0.5truein}
\begin{center}
\includegraphics[scale=0.08]{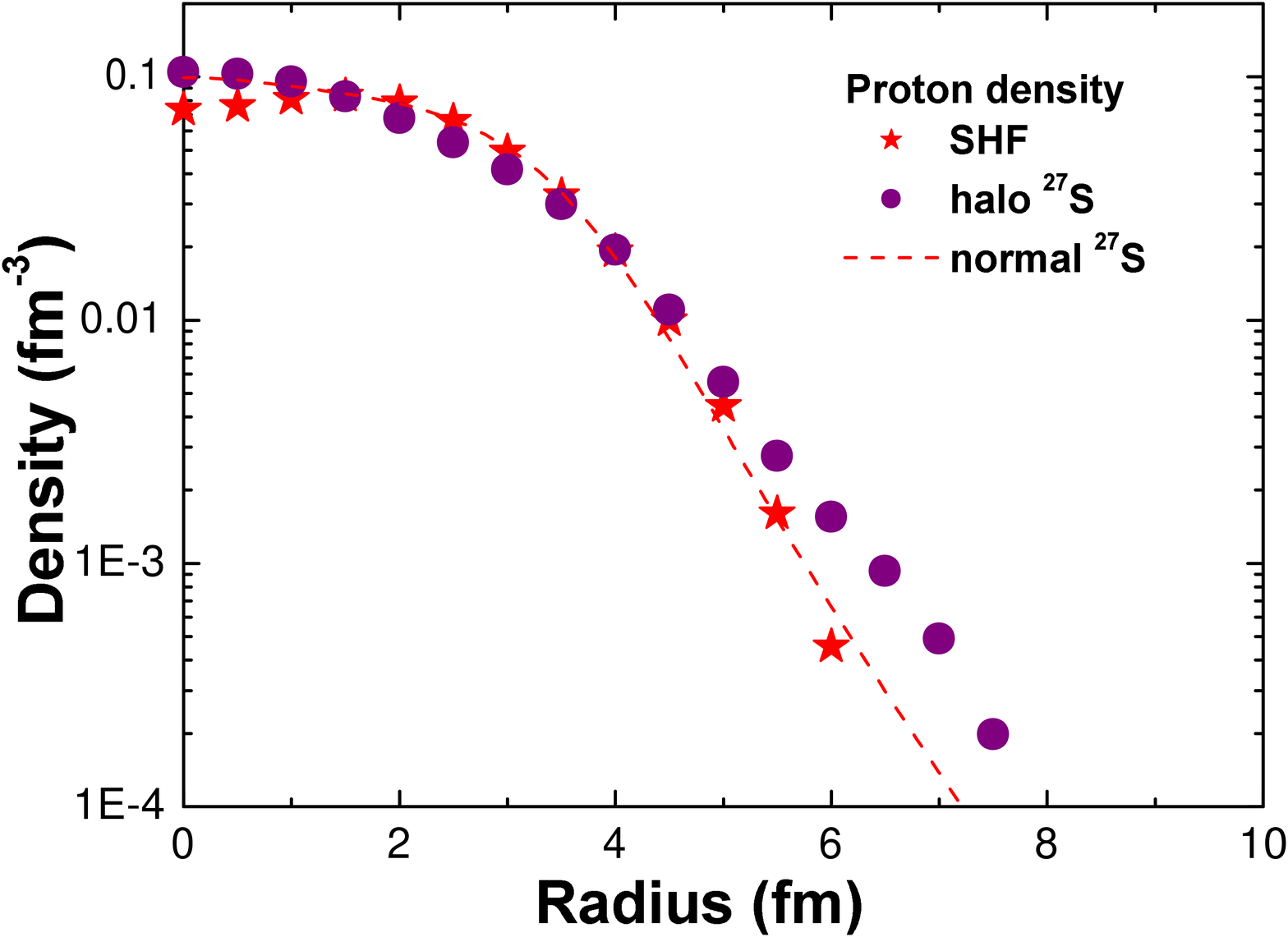}
\includegraphics[scale=0.08]{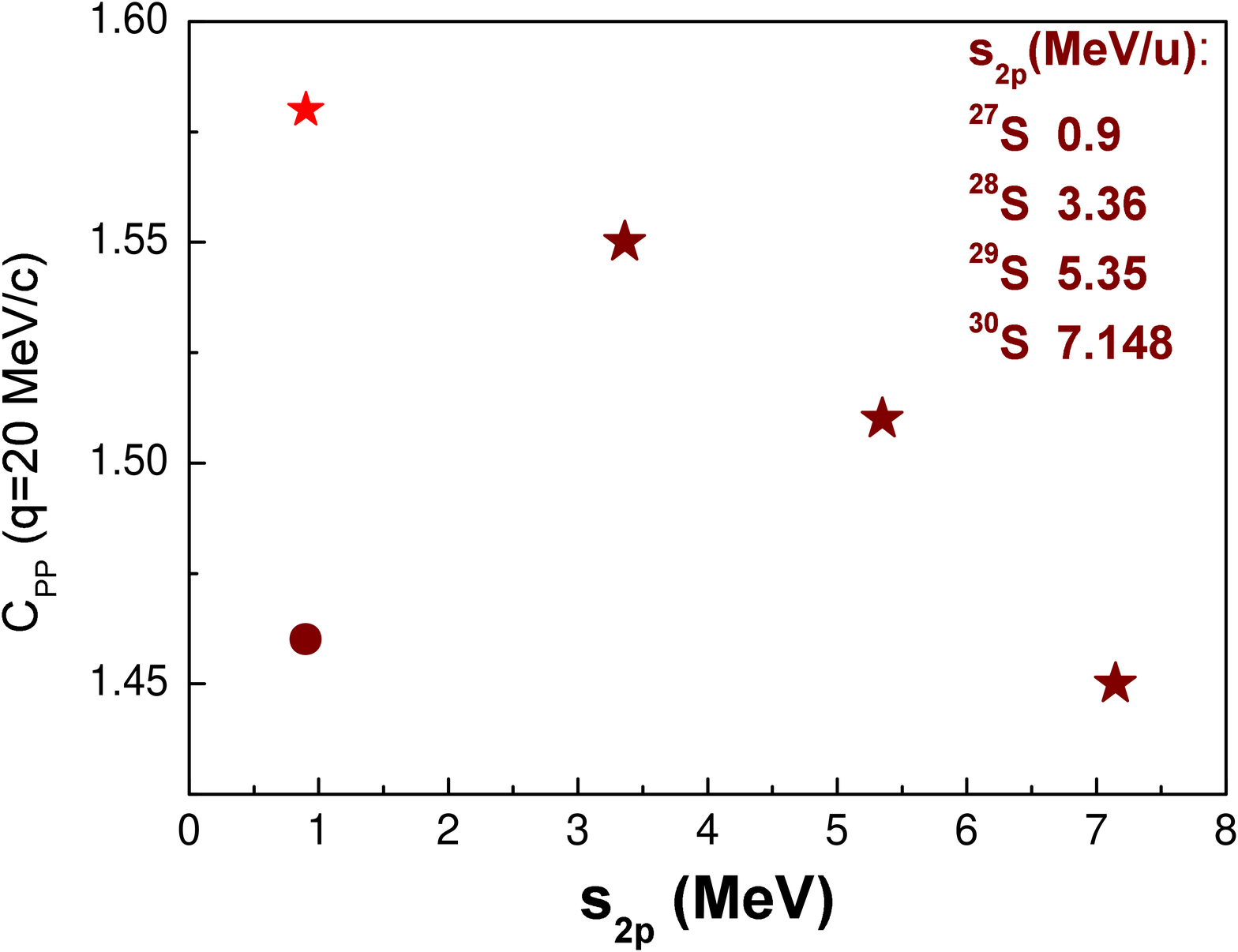}
\end{center}
\vspace{-0.6truein} \caption{ The left panel: the density
distribution of protons as a different initial input in IDQMD.
Right panel: p-p correlation function strength at 20 MeV/c as a
function of the two-proton separation energy. See details in text.
\label{Fig3} }
\end{figure}

After we discussed neutron-neutron, neutron-proton correlation
functions, we shall investigate proton-proton correlation
function. To this end, we selected some proton-rich nuclei,
namely, $^{27,28,29,30}S$. Since $^{27}S$ was predicted as
two-proton halo nucleus, we would test the sensitivity of
proton-proton  correlation function to the proton density
distribution. The left panel of Fig. 3 gives the density
distribution of protons which were used as a different initial
input in IDQMD calculation. The stars and the line show the proton
density distribution with the usual Skyrme-Hartree-Fock
calculation and the normal IDQMD initialization, respectively. In
these cases, no special consideration for proton-halo density
distribution. While the circles represent a halo-proton density
distribution which is characterized by its long density tail. Now
we compare the proton-proton correlation function ($C_{PP}$) with
either the normal density distribution (star) or the proton-halo
density distribution (circle). The right panel of Fig. 3 shows the
p-p correlation function strength at 20 MeV/c as a function of the
two-proton separation energy ($s_{2p}$) of $S$-isotopes. If there
is no special proton density distribution, the p-p correlation
strength decreases with $s_{2p}$ or the mass number of
$S$-isotopes, which just illustrates that the system tends to be
more loose when the mass number of $S$-isotopes increases.
However, $C_{PP}$ dramatically decreases when the proton-halo
density is assumed in the QMD initialization (see circles in the
figure). This kind of decreasing essentially reflects the extended
density distribution of proton and incompact system of $^{27}S$.

In summary, nucleon-nucleon momentum correlation functions from
light nuclei induced reactions have been systematically
investigated and its sensitivity to the binding energy or
separation energy of weakly-bound nuclei has been explored from
the break-up reactions of nuclei in the framework of the IDQMD
model. Firstly we gave a well-fitted halo neutron - halo neutron
correlation function from the break-up of $^{11}Li$ on $C$ target.
Based upon this achievement of the good fit, we explore the
dependence of the proton-neutron correlation function ($C_{PN}$)
at small relative momentum with the binding energy  for $Li$, $C$
and $N$ isotopes. It was found that the correlation strength of
$C_{PN}$ at small relative momentum rises with the the binding
energy. This changeable tendency of $C_{PN}$ with $E_{b}$  might
be a potential good way to study the spatial structure of the
light nuclei. In addition, the proton-proton correlation function
shows a dramatic decrease for the proton-halo density
distribution, which provides us another potential tool to diagnose
proton halo nuclei.

This work was supported in part by the Shanghai Development
Foundation for Science and Technology under Grant No. 05XD14021,
06JC14082, the National Natural Science Foundation of China under
Grant No. 10535010, 10328259 and 10135030.

{}

\end{document}